# Photoconductive response of strained silicon nanowires: A Monte Carlo study

Daryoush Shiri[1], Amit Verma[2*], Mahmoud M. Khader[3]

[1]Institute for Quantum Computing (IQC), Department of Physics and Astronomy, University of Waterloo, Waterloo, N2L3G1, Ontario, Canada
[2]Department of Electrical Engineering, Texas A&M University-Kingsville, Kingsville, Texas, 78363, USA
[3]Gas Processing Center, Department of Chemistry, Qatar University, P. O. Box 2713, Doha, Qatar

[*]Corresponding Author: Amit.Verma@tamuk.edu.

Using Ensemble Monte Carlo simulations the photocurrent in a 500nm long strained [110] silicon nanowire with diameter of 3.1nm is investigated. It was observed that a phototransistor based on this nanowire can have responsivities in the order of 21.3 mA/W for an input light wavelength of 532 nm and intensity of 0.25-2.5 kW/cm$^2$. The super-unity slope of 1.2 in photo conductance versus input light intensity suggests that the nanowire has a photoconductive gain and highlights its advantage over germanium nanowires with sub-unity slope (0.77). The generated photocurrents are in the 0.1 nA-1 nA range. Density Functional Theory (DFT) and Tight Binding (TB) methods were used for strain application and band structure calculation, respectively. Both longitudinal acoustic and optical phonons were included in the calculation of the carrier-phonon scattering events, which showed a two-order of magnitude stronger role for longitudinal optical phonons.

## I. INTRODUCTION

There has been intensive research on the photoluminescence (PL) properties of silicon nanowires (SiNWs) over the last few years, and now the evidences of quantum confinement, direct bandgap, and strain induced changes of PL peak in SiNWs have been reported [1-2]. Fabrication of narrow SiNWs with diameters of 3nm or less is now possible using both top-down [3] and bottom up [1-2] methods through self-limited oxide assisted narrowing. The sources of strain in SiNWs are also abundant either intrinsically (e.g. due to lattice mismatch with cladding material) [2][3] or extrinsically (e.g. applied by deformable substrates) [4]. It was shown that strain can change the spontaneous emission time in SiNWs by a few orders of magnitude [5] through an already observed change in the band structure from direct to indirect [6-9]. This significant increase in recombination lifetime (implying a reduction in the recombination rate) in particular makes small diameter strained SiNWs potentially very attractive as sensitive photodetectors. Also the radial strain induced from the silica matrix or SiO$_2$ cladding can result in the shift of PL peak (through the change of bandgap) [1-2]. This was further proved by further Tight Binding (TB) modeling of radial strain effect on bandstructure in [1]. On the other hand SiNWs are promising for applications in phototransistors [10] and avalanche photodiodes [11]. The silicon nanowires in the aforementioned applications [10-11] have large diameters and as a result they have indirect bandgap as bulk silicon does. However the optical anisotropy due to local field effects and better control of electrostatic of the device via gate-all-around (GAA) topology makes them advantageous.

The main scope of this work is studying the dynamic of carriers in a strained narrow [110] SiNW in response to time dependent photo excitation. The Ensemble Monte Carlo (EMC) simulation is used to study the effects of multi-phonon carrier scattering events on the photo-generated electrons and holes under the influence of electric field and temperature. The electric current induced by photo excitation is calculated and its dependence on the incoming photon flux and electric potential is investigated. For incoming light intensities of 0.25 kW/cm$^2$ and 2.5 kW/cm$^2$, the average induced photo current is 26 pA and 0.8 nA, respectively (for V$_{DS}$ = 2 V). In section II the computational methods, simulation settings and assumptions are introduced. The third section is devoted to the computational results and discussions followed by conclusion in section IV.

## II. METHODS

### A. Simulation setup

**A.1 Band structure:** The structural relaxation of a nanowire and application of strain were done with DFT method implemented in SIESTA package [12]. The band structure and Eigen states were calculated with 10 orbital $(sp^3d^5s^*)$ TB method using the Jancu's parameters as given in [13]. This is to avoid the inherent bandgap underestimation of DFT and diameter sensitive many–body GW corrections. This TB method has successfully reproduced the experimental data regarding the effect of radial strain on the PL spectrum of narrow SiNWs [1-2]. The trend of TB bandgap change with diameter in the case of silicon nano-crystals also agrees with DFT based calculations [14]. In the case of [110] SiNWs, TB method can reproduce the bandgap trend with diameter as observed in Scanning Tunneling Spectroscopy (STS) measurements [15]. In addition the TB method has been successful in regenerating the bulk silicon band structure as well as correctly simulating the boundary conditions i.e. surface passivation. Sun et al observed that intrinsic +2% strain in Ge-on-Si light emitting diodes results in direct bandgap for germanium [16]. We further confirmed this value by performing a TB calculation of strain effects on bulk germanium. The nanowire in this work was cut from bulk silicon crystal in [110] direction and surface terminated with hydrogen atoms. The cross section of nanowire lies in the x-y plane and z is the axial direction of each nanowire. Figure 1 shows the atomic structure of a unit cell in a SiNW oriented in [110] direction with a diameter of 1.7 nm.

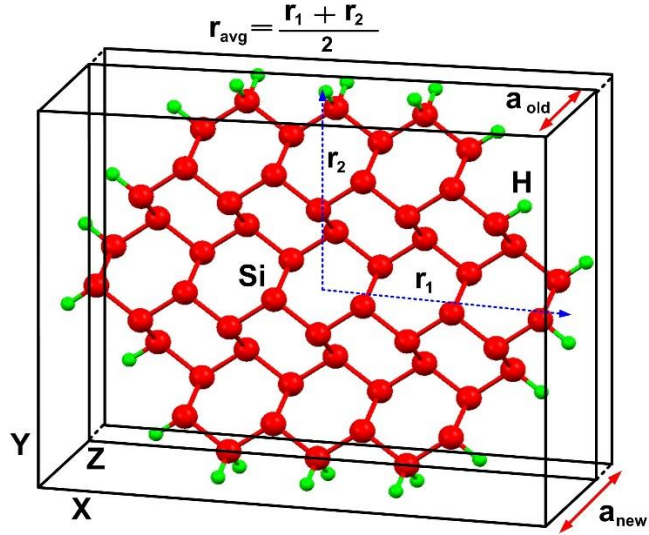

FIG. 1. Illustration of a [110] SiNW unit cell with diameter of 1.7nm. The diameter is defined as the average of large and small diameters. Dark and bright atoms correspond to Si and H, respectively. Application of strain is performed by updating the old unit cell ($a_{old}$) by the given strain percent ($\varepsilon$).

The relative stability of [110] direction compared to [100], which is quantified as free energy of formation [17], is experimentally verified in [15][18]. Energy minimization of nanowires is performed by DFT method within SIESTA [8] package using Local Density Approximation (LDA) functional with Perdew-Wang (PW91) exchange correlation potential [19]. Spin polarized Kohn-Sham orbitals of double-$\xi$ type with polarization (DZP) were used. The Brillion Zone (BZ) was sampled by a set of $1 \times 1 \times 40$ k points along the axis of the nanowire (z axis). The minimum center to center distance of SiNWs is 6nm to avoid interaction between nanowire unit cells due to wave function overlapping. Energy cut-off, split norm and force tolerance are 680 eV (50 Ry), 0.15, and 0.01 eV/Å, respectively. The energy of unstrained unit cell of nanowire is minimized using Conjugate Gradient (CG) algorithm during which the variable unit cell option is selected i.e. the unit cell length of nanowire can relax to a new value according to the given force tolerance as above. At this stage the unstrained energy minimized unit cell length is updated i.e. $a_{new} = a_{old}(1+\varepsilon)$, where a is unit cell length and $\varepsilon$ is strain value in percent (see Figure 1). The new unit cell is then relaxed using the constant volume (fixed unit cell) option in which the atoms are only free to move within the fixed unit cell volume. The result of each minimization step is fed to the next step of minimization to increase or decrease the strain depending on its tensile or compressive nature. The nanowire in this study obtains an indirect bandgap due to applied axial strain of -5%. The conduction band minimum at BZ center ($k_z=0$) is 80 meV above the indirect conduction band minimum ($E_{cmin}$ in Figure 2.a). In performing the EMC simulation, tabulated values of two lowest conduction sub bands ($C_1$ and $C_2$ in Figure 2.a) and four highest valence sub bands ($V_1$ to $V_4$ in Figure 2.a) are used. This is because the rest of the sub bands are at least 100 meV apart from the chosen sub bands within the band structure. It is therefore expected to have negligible charge carrier occupation even at relatively high electric fields, similar to what has been observed in unstrained SiNWs of similar diameters [20].

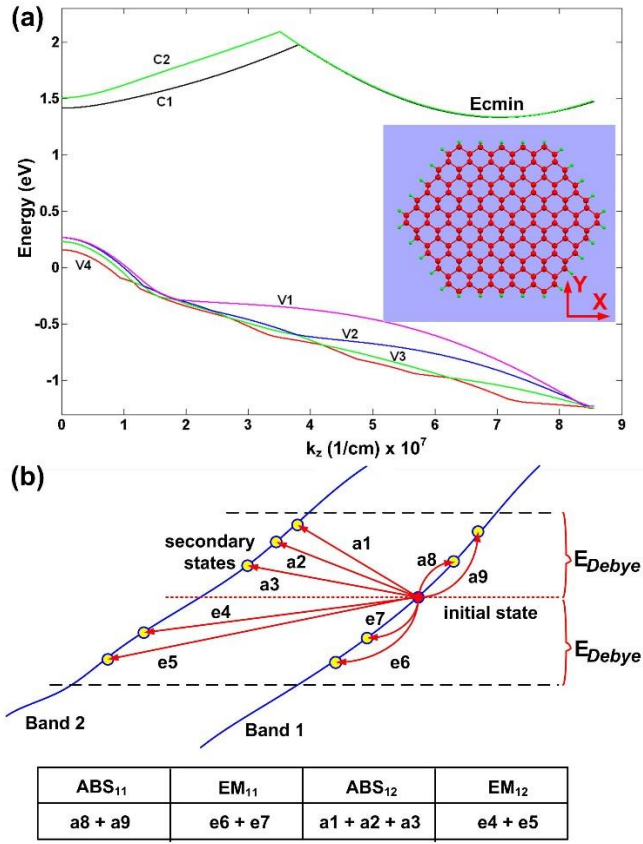

| ABS$_{11}$ | EM$_{11}$ | ABS$_{12}$ | EM$_{12}$ |
|---|---|---|---|
| a8 + a9 | e6 + e7 | a1 + a2 + a3 | e4 + e5 |

FIG. 2. (a) Band structure of a 3.1 nm [110] SiNW showing indirect bandgap with band offset of 80 meV. Two conduction (C) and four valence (V) bands are used in this simulation. Inset shows the cross section of the nanowire in *xy* plane. (b) Grouping and sorting individual LA phonon absorption (a) and emission (e) rates according to intra- or inter-sub band nature of the event. ABS and EM represent absorption and emission of phonons, respectively. A subscript like 12 means scattering from band1 to band2. The Monte Carlo code uses individual scattering rates hence the plus sign (+) means grouping the scattering events in this context. To calculate the total scattering rates however the grouped rates are added together as shown by + sign.

**A.2 Scattering Rates:** The electron/hole-phonon scattering rates were calculated using first order perturbation theory [21]. Generally the total scattering rate corresponding to a given state in BZ e.g. $k_z$ can be written as a summation over individual scattering rates i.e.:

$$W_{k_z} = \sum_{k_z'} W(k_z, k_z'). \qquad (1)$$

Where $W(k_z, k_{z'})$ stands for scattering rate of the carrier from an initial state at $k_z$ to all possible final states (at $k_{z'}$). Both inter- and intra-sub band electron-phonon scattering events were taken into account in $W(k_z, k_{z'})$. For each initial state starting from a specific sub band (C for electrons and V for holes), all possible final states are found according to the phonon type. In the case of Longitudinal Acoustic (LA) phonons all states which are within a window of $E_z \pm E_{Debye}$ (starting from initial state energy, $E_z$), constitute all possible $k_{z'}$. $E_{Debye} = 63$ meV is the Debye energy or the maximum energy of LA phonons. Each individual carrier-LA phonon scattering rate, i.e. $W(k_z, k_{z'})$, is given as:

$$W(k_z, k_z') = \frac{D_{e/h}^2}{8\pi^2 \rho \hbar^3 v_s^4} \Delta E_{kk'}{}^2 B_{\pm}\left(\left|\pm \frac{\Delta E_{kk'}}{\hbar v_s}\right|\right) \Phi(q_t, q_z). \qquad (2)$$

Where $D_{e/h}$, $\rho$ and $\upsilon_s$ represent electron (hole) deformation potential ($D_e = 9.5$ eV and $D_h = 5$ eV), mass density ($\rho = 2329$ Kg/m$^3$) and velocity of sound in silicon ($\upsilon_s = 9.01 \times 10^5$ cm/sec).

B and $\Phi$ are Bose-Einstein and structural factors, respectively and are defined in [22]. $\Delta E_{kk'}$ is the energy difference between initial and final states and $q_t$ and $q_z$ are transverse and longitudinal wave vectors of phonons. In case of carrier scattering due to Longitudinal Optical (LO) phonons, the secondary states are those which are exactly $E_{LO}$ above or below the energy of initial state (at $k_z$). $E_{LO} = 54$ meV is the maximum energy of LO phonons by assuming a dispersion-less branch of LO phonons. The individual rate, i.e. $W(k_z, k_{z'})$, in this case is written as:

$$W(k_z, k_z') = \frac{|D_{op}|^2}{8\pi^2 \rho \omega_0} \int_0^{q_c} \Phi(q_t, q_z) dq_t \, B_{\pm}(E_{LO} = \hbar \omega_0) \frac{1}{\left|\frac{\partial E(k_z)}{\partial q_z}\right|_{k_z = k_p}}. \qquad (3)$$

Where $D_{op}$ is deformation potential of LO phonons for electrons ($D_{op} = 11 \times 10^8$ eV/cm) and holes ($D_{op} = 13.24 \times 10^8$ eV/cm). $\omega$o is the maximum phonon frequency of LO phonons. $q_c$ is the maximum allowable value of phonon transversal component ($q_t$) within the BZ of bulk silicon which is $q_c = 1.9\pi/a$ [22]. The last term in equation 3 is density of states evaluated at each available final state ($k_p$) into which carrier will scatter. $q_z$ and $k_z$ are wave vectors of phonons and electrons, respectively. They step together due to momentum conservation in the summations involved [22].

The carrier-phonon interaction Hamiltonians in previous equations are of deformation potential type for bulk LA and LO phonons. Detailed procedure of deriving equations 2 and 3 from first order perturbation theory using Deformation potential Hamiltonian were explained in Appendix. I. In unstrained small diameter SiNWs, it was shown to be a reasonable approximation to confined phonons and allows for an evaluation of inter sub band charge-carrier scattering in a relatively straightforward way [5]. We assume that this will also be the case for similar diameter SiNWs with relatively small strain. Also since the electron-phonon scattering process is many orders of magnitude faster (~ psec) than the electron-hole recombination times (~ μsec - msec),  we do not expect that a small change in the total carrier-phonon scattering due to bulk and confined phonon difference will affect the quantitative and qualitative nature of our results.

Standard EMC methodology [21] is applied in developing and performing the simulation to investigate drift and effects of multi-phonon carrier scattering events on photo-generated electrons and holes under the influence of applied voltage. Within our EMC simulation setup, electron and hole transport is confined to the first BZ, which is divided into 8001 $k_z$ grid points (i.e. 4000 positive $k_z$ and 4000 negative $k_z$ values) and for which the tabulated energy values and carrier-phonon scattering rates, as well as the possible final states after scattering, are computed and fed into the EMC code. Figure 2.b shows how the grouping of individual phonon emission/absorption rates and adding them together is done for electron-LA phonon scattering events. The calculated rates are then sorted and stored according to inter- or intra-sub band scattering type. Decision is made based on the number of $k_{z'}$ (secondary state indices) which determines if the secondary state belongs to *band1* (intra-sub band) or *band2* (inter sub band). It is noteworthy that this is a simplified picture for the sake of visibility since in the case of LA phonons there are indeed many secondary states similar to a quasi-continuum within the Debye energy window.

**A.3 Electrostatics:** The photo detector device in the simulation is composed of a 500 nm long intrinsic (undoped) SiNW with 3.1nm diameter and [110] crystallographic axial direction. It is biased between two drain and source terminals (see Figure 3). It is assumed that the nanowire is surrounded by air and the gate voltage at a point 10 micrometers away (i.e. a large distance away) from the nanowire is $V_G$=0. So essentially the electric field from the nanowire into the surrounding is small. Also this gate-induced electric field is perpendicular to the nanowire peripheral surface. This results in a constant electric potential within each cross section of the nanowire. Such a structure of a strained SiNW (with surface states terminated by hydrogen) is used to investigate a lower bound on the performance. Depending upon the shape and structure of the gate as well as the gate and insulating materials, actual grounded gate nanowires will most likely result in faster decay time and help in higher frequency operations. The nanowire is assumed to be uniformly at room temperature (300K).

Device electrostatics is evaluated by solving the Gauss Law in integral form self-consistently with EMC [23]. Finding the potential profile in the SiNW photo detector requires solving 3D Poisson and 1D transport problem together. The difficulty that arises due to this combination can be avoided by using Gauss Law. Since each cross section of the nanowire is assumed to be equipotential, the nanowire in this work is considered to be composed of a train of Gaussian surfaces enclosing each grid point for which the potential is to be determined. The grid points are considered to be 0.5 nm apart. Two of the surfaces (each 3.1nm × 3.1 nm) are perpendicular to the nanowire axis, while the other four correspond to SiNW-air boundary (each 3.1nm × 0.5 nm). Therefore the closed surface integral in Gauss Law becomes a summation of 6 surface integrals as given below:

$$\epsilon \oint_{area} \overline{E}.\overline{ds} = Q_{total} = eL\big(p_i + N_D^+ - (n_i + N_A^-)\big). \qquad (4)$$

Where ($Q_{total}$) is the total charge enclosed in that element which is obtained from EMC simulation. E is the electric field and L is the grid spacing (0.5 nm). $p_i$, $n_i$, $N_D$, $N_A$ correspond to hole, electron, donor and acceptor densities, respectively.

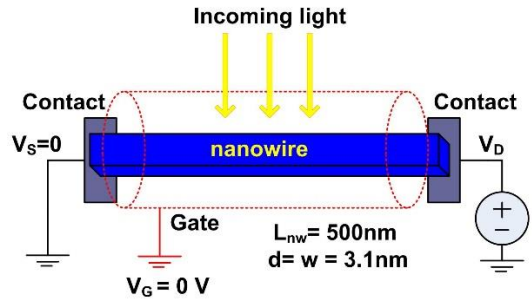

FIG. 3. Schematic of SiNW photo detector and electrostatic parameters assumed in EMC simulation.

The nanowire is also assumed to be in thermal equilibrium with the contacts corresponding to the Ohmic contacts assumption. The electrons and holes that are generated due to photon influx drift under the influence of the longitudinal electric field due to the applied bias.

B. **Optical Parameters**

The photon flux is assumed to be uniformly distributed along the length of the nanowire. It was previously reported that the diameter of diffraction limited laser spot in photocurrent and PL spectroscopic studies of silicon [10] and germanium nanowires [24] is about 500 nm. ADDED: For simplicity we assume that the photon fux is uniform along the length of the nanowire which is 500nm. The electric field component of the incoming photon field is parallel with the length of the nanowire (z axis). It has been experimentally verified that the absorption or emission of this component is dominant due to the strong local field effect [25]. To be able to compare the results with existing experimental data, the chosen wavelength is λ=532 nm (green light), which corresponds to the photon energy of 2.33 eV. Since this nanowire is of indirect bandgap we use the absorption spectrum of bulk silicon. Note that this may not apply to the case of direct bandgap nanowires. At the given photon energy of 2.33 eV, the value of absorption is 10,000 $cm^{-1}$, which corresponds to an absorption depth of 1μm as it was also observed for indirect bandgap nanowires in [24]. Since the calculated photocurrent can be scaled by quantum efficiency (η), it is assumed that η=1. However due to nonzero reflectivity (R=0.4) of silicon at the given wavelength (532 nm) and inefficient absorption of photons due to the thin nanowire (d=3.1 nm), the total quantum efficiency is reduced to $(1-R)(1-e^{-\alpha d}) = 0.186\%$.

The quantity that relates the incoming light intensity ($P_{in}$) to the rate of electron-hole generation in nanowire is the photon flux (Φ), which is defined as $\Phi = P_{in}A/\hbar\omega$. The photon collecting area (A) is 3.1 nm × 500 nm. Therefore for input light intensities of 0.25 $kW/cm^2$ and 2.5 $kW/cm^2$, the photon flux is $10^9$ $s^{-1}$ and $10^8$ $s^{-1}$, respectively. This means that on average during a 0.1 nsec period, at each 0.5 nm of the nanowire length (which was the grid size we considered in our EMC simulations), 0.01 and 0.001 pairs of electron-hole are generated, respectively. These numbers are compared with a random number uniformly generated between 0 and 1 within the EMC simulation to determine the generation of an electron-hole pair at each grid point at the start of every 0.1 nsec period for which the nanowire is illuminated. The input intensity lies within the range of experimental values of 0.1 $kW/cm^2$-10 $kW/cm^2$ [24]. In principle assuming any value for the on-off rate of the incoming light pulse is possible in this simulation. However low frequency input light pulse places additional memory requirement. As an example if $T_{pulse}$=1 ms and the number of absorbed photons is $1\times10^8$ $s^{-1}$ (as reported experimentally in [24]), then during the 1 ms pulse duration the number of generated carriers is $1\times10^5$. However studying the time evolution of $10^5$ electron-hole pair under the influence of electric field and scattering events requires large amount of memory. Limiting the light intensity pulse period to 10 ns means that at the very least one electron-hole pair is captured and followed by the simulator. Provided that memory and computing power allows us to calculate the evolution of more than 105 electron-hole pairs, it is possible to simulate the photocurrent for longer pulse durations. In this case the values of photocurrents scale up. In addition to the scattering events due to phonons, radiative recombination of electron-hole pair is another process that can potentially reduce the photocurrent value. Since the nanowire in this study is of indirect bandgap, the recombination process is a slow second order process mediated by phonons. The room temperature radiative recombination time for bulk silicon is about $\tau_{rad}$=100 μs. As it will be seen, this is much greater than the time scales of carrier-phonon scattering rates as well as input light modulation speed. For SiNWs of similar diameters it was shown that the radiative recombination time in indirect bandgap nanowires can be orders of magnitude larger than microseconds [5]. Therefore the recombination process was ignored in the EMC simulation unless we deal light pulses which are as long as electron-hole recombination time e.g. >1msec in the nanowire of this work.

**Results**

### C. **Electron (hole)-phonon scattering rates**

The electron-phonon scattering rates for initial states in the positive half of the BZ are shown in Figure 4.a. Here the total rate, i.e. the summation of inter sub band and intra sub band scattering rates are shown for these scattering events in which the initial states reside in the first or second conduction sub band ($C_1$ or $C_2$). As can be seen, the role of LO phonons is two orders of magnitude stronger than the scattering events mediated by LA phonons. This is a feature similar across many nano materials such as single-wall carbon nanotubes (CNT) [26] and unstrained SiNWs [20].

Even the slowest electron-LA phonon scattering times (~1 ps) justify ignoring the recombination time in our simulation. The stronger role of LO phonons is also evident in Figure 4b which shows the hole-phonon scattering rates for the scattering events starting from the first valence band ($V_1$). Compared to the electron-LO phonon scattering, the hole-LO scattering events are faster due to larger deformation potential for valence sub bands. Note that the sharp peaks in LO-phonon induced scattering rates are due to van Hove singularities that occur in 1D density of states i.e. last term in equation 3. Also the occurrence of the first peak in carrier-LO phonon scattering reveals the first state in BZ (starting from $k_z$=0) after which phonon emission becomes possible. In other words this is the first point in which the energy of electrons (holes) becomes higher (lower) than E at $k_z$=0 by $E_{LO}$.

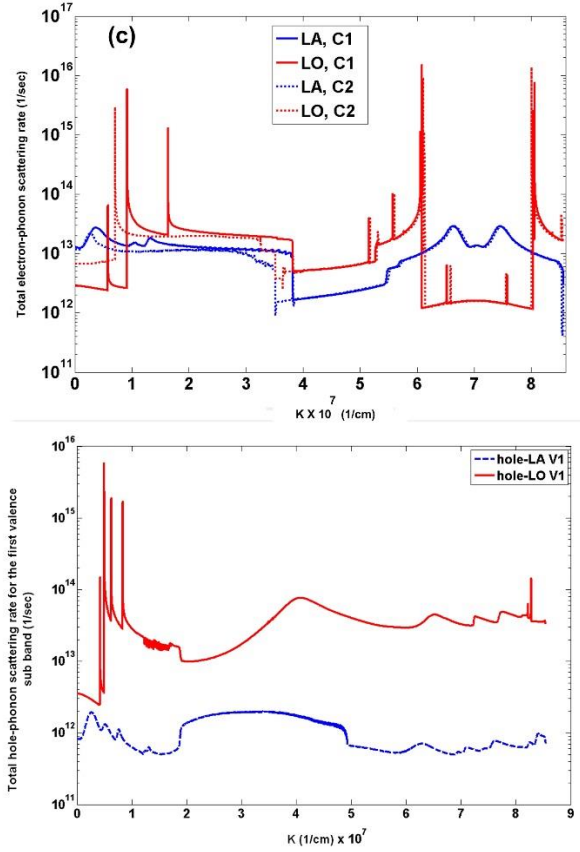

FIG. 4. (a) Electron-phonon scattering rates in a strained 3.1 nm [110] SiNW. $C_1$ or $C_2$ stand for the sub bands in which the first state resides before being scattered by phonons. (b) Hole-phonon scattering rate for the first valence sub band ($V_1$) of the same nanowire. The dominant role of LO phonons is evident in both.

## D. **Photocurrent**

Figure 5 shows the photo response of the SiNW FET that is biased at $V_G = 0$ V and $V_{DS} = 2$ V and illuminated by the maximum intensity of 2.5 kW/cm². The light pulse duration is 10 ns and the simulator samples the electron and hole current each 1 ns. By decreasing the light intensity at a constant bias voltage, the number of generated electron-hole pairs is reduced, which results in smaller photocurrent values. It should be noted that the noise observed in the photocurrent in Figure 5 is intrinsic and emanates from the small number of electrons and holes collected during each sample period. This number is randomly different from one sample period to another.

As can be seen in Figure 6a the current at 0.25 kW/cm² is lower by one order of magnitude. It is also evident that during the light pulse duration there are some time samples at which the current is zero. Increasing the electric field along the nanowire length i.e. increasing $V_{DS}$ means that the carriers have more chance to arrive at the terminals. This is because we are still below the velocity saturation regime. Therefore increasing the electric field reduces the transit time, $\tau_{transit}$, and makes it comparable or less that scattering time due to phonons ($\sim$ psec). This can also be observed in Figure 6b that compares the total photo current at two different bias voltages ($V_{DS} = 1V$ and $V_{DS} = 2V$) for a constant maximum light intensity of $P_{in}$=2.5 kW/cm$^2$.

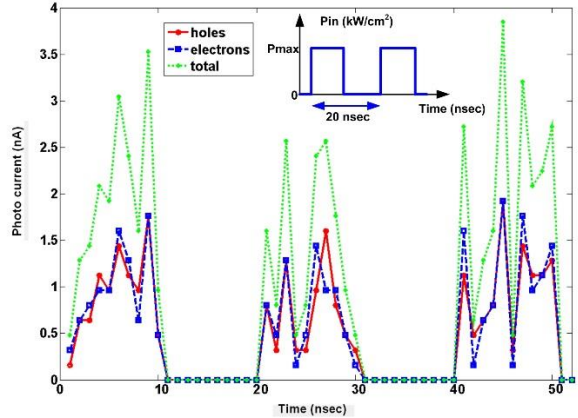

FIG. 5. Photocurrent induced in the SiNW in response to the input laser pulse (Inset) with the frequency of 500 MHz. The total current is composed of the current due to electrons (holes) which arrive at drain (source) terminal.

To obtain a better picture of the relevance between input light intensity and photocurrent, the average photocurrent of SiNW FET is calculated for a constant duration of 100 ns for different bias voltages. Figure 7a shows the average photocurrent versus the input light intensity. As can be observed at two different voltages the difference between the photocurrents is more pronounced when the input light intensity is high. It is instructive at this point to compare the responsivity of the SiNW-based photo transistor with that of a CNT-based photo transistor. According to Figure 7a, the responsivity of the SiNW FET for 2.5 kW/cm$^2$ and 0.25 kW/cm$^2$ input intensities (at $V_{DS} = 2V$) is 21.3 mA/W and 0.7 mA/W, respectively. On the other hand the responsivity of an array of 100 μm long and 1-2nm thick CNT-FET is reported as 9.87×10$^{-5}$ A/W [27]. This strongly suggests that by fabricating an array of SiNWs the responsivity can be increased and a better performance compared to CNT-based photo-detectors can potentially be achieved. Recalling that photocurrent is proportional to $\tau_{recomb}/\tau_{transit}$, it is evident that for a comparable length, bias voltage and photon flux, CNTs generate much smaller photocurrent because of their lower $\tau_{recomb}$ values. PL study of semiconducting CNTs has shown very fast recombination life times in the order of 30 psec [28-29]. On the other hand the recombination lifetime in direct bandgap SiNWs is in the order of microseconds [1-2]. Also our computational studies suggest [5] that in indirect bandgap SiNWs the recombination lifetime can be as slow as a few milliseconds or more.

It is important, however, to point out that the SiNW in our model is ideal i.e. the surface effects were obviated by Hydrogen passivation. However the CNT photo-detectors reported in [27] were encased in a PMMA polymer layer. Therefore in addition to the difference in recombination lifetimes, the lower reported photocurrent in CNT photo-detectors compared with SiNW may also be attributed to the extra absorption due to PMMA layer, or the surface effects which merits more DFT study. In more recent studies like [30] and [31] it was also shown that addition of PMMA or DNA layers on CNT can reduce the light absorption. However based on the observed results it is surmised that including realistic experimental effects e.g. surface effect in arrays of SiNW photo detectors will at its worst makes SiNWs on par with that of CNT.

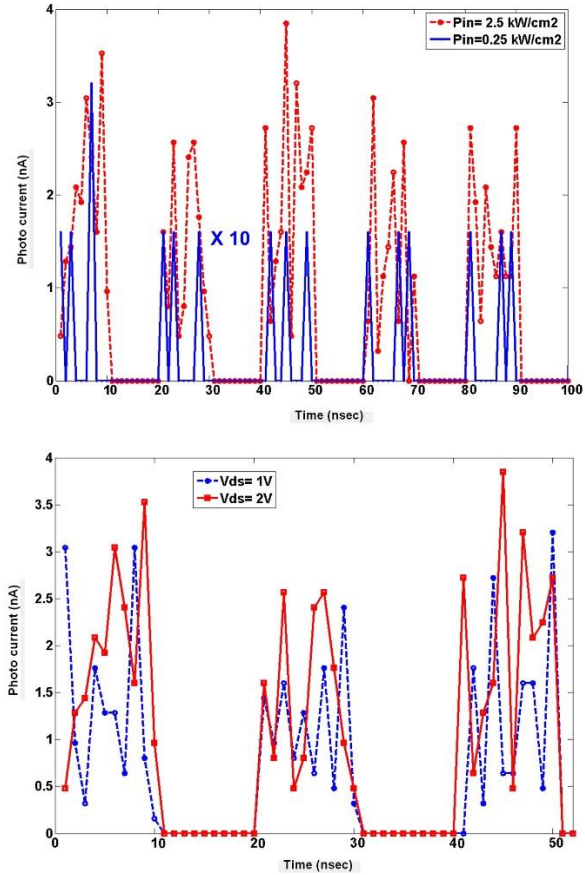

FIG. 6. (a) Photocurrent induced in the SiNW in response to two different laser input powers at a constant drain-source voltage ($V_{DS}$=2 V). The case of $P_{in}$=0.25 kW/cm$^2$ is magnified by a factor of 10 for the sake of visibility. (b) Photocurrent at constant input power of 2.5 kW/cm$^2$ at two different applied drain-source voltages.

Figure 7b shows that by increasing the input intensity the photo conductance of the SiNW is enhanced significantly. For input light intensities of 0.25 kW/cm$^2$ and 2.5 kW/cm$^2$ the values of photoconductivity which is defined as G = $\Delta$I/$\Delta$V (slope of the lines), are $G_1$= $\Delta I_1/\Delta V$ = 1×10$^{-11}$ S and $G_2$= $\Delta I_2/\Delta V$ = 2×10$^{-10}$ S, respectively. This implies that by applying a train of light pulses the conductivity of SiNW can change by 0.2 nS, provided that we assign a conductivity of zero value to dark current (off) case. Although this looks inferior compared to the reported conductivity of 2.5 µS in germanium nanowire FET [24], the difference can be attributed to the large difference of diameters as well as input light power in that reported work compared to our work. In the experiment with germanium nanowire the diameter is 50 nm and the input light intensity is 10 kW/cm$^2$.

In contrast to germanium nanowires, which show a sub-unity exponent (slope=0.77), the rate of change of photo conductivity with the input intensity in SiNW is 1.2, which is the manifestation of a photoconductive gain. This means that electron and holes have enough time to rotate in the circuitry (combined nanowire and voltage source) many times before the next recombination event occurs. However this slope is different from the experimental value of 1.00 reported for SiNWs [24]. This discrepancy can be attributed to the difference between the idealized nanowires in our simulation (e.g. smaller diameter and surface termination with hydrogen) and the experimental nanowires which have larger diameters and possibility of surface recombination due to non-ideal surfaces. Larger surface as a result of larger diameter can lead to more pronounced surface effects as well. In spite of this difference, it is clear that given the reported experimental results on SiNW, GeNW and CNTs, as well as our theoretical results, SiNWs have the potential for photoconductive gain. By the advent of new techniques to diminish surface effects [1-3], this may not be far from reach.

### III.  CONCLUSIONS

The performance of a strained SiNW FET photo detector was computationally investigated using Ensemble Monte Carlo simulations. Electron and hole transport in the simulation was LA and LO phonon limited, and inter- and intra-sub band scatterings events were included. The photo response of the SiNW to a 500 MHz input light pulse was calculated. The induced currents are in the range of 0.1 nA-1 nA and the change of conductance is on the order of 0.1 nS. It was shown that with input light intensity of 2.5 kW/cm$^2$ and the bias voltage of V$_{DS}$=2 V, the responsivity of 21.3 mA/W can be achieved. This is two orders of magnitude more than the values reported for arrays of CNT-FET devices of comparable topology. Comparison of photoconductivity change by input light intensity suggests that SiNWs can potentially exhibit super-unity slope i.e. photo-conductive gain as opposed to germanium nanowires. Availability of strain application methods to SiNWs e.g. using deformable substrates can also add a new degree of freedom to tune the photo generated current in SiNWs through the variation of band structure and absorption. Given that the compatibility with the mainstream silicon technology makes the implementation of SiNW-based FET arrays more feasible than CNT-based counterparts, the results further highlight the potential of small diameter SiNW based photo-detectors.

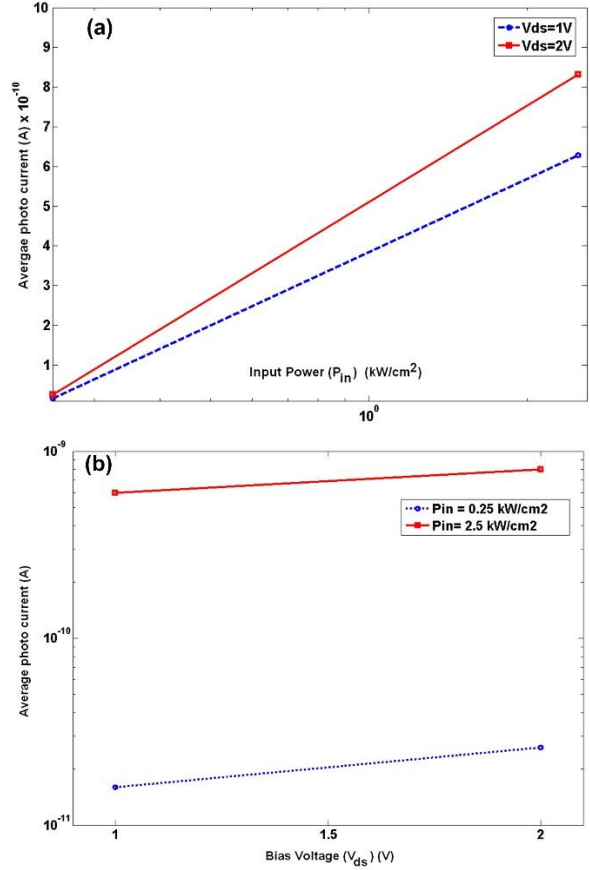

FIG. 7. (a) Average photocurrent vs. input power showing gain modification by increasing the applied voltage. (b) Average value of photo current vs. drain-source voltage at two different input intensities reveals how much photo conductance is induced in each case.

## APPENDIX A: CALCULATION OF ELECTRON (HOLE)-PHONON SCATTERING RATES

Figure A.1 shows a simple example of inter-sub band scattering in which an electron at the bottom of the indirect conduction band is scattered into many available secondary states within $E_{Debye}$ window. If the rate of each scattering event is called $W(k_z, k_{z'}, \mathbf{q})$, then the total scattering rate of the electron at $k_z$ is found by summation over all available secondary states ($k_{z'}$) and phonon wave vectors [$\mathbf{q}=(q_x, q_y, q_z)=(q_t\cos\varphi, q_t\sin\varphi, q_z)$] i.e.:

$$W_{k_z} = \sum_{k_z', \widetilde{q}} W(k_z, k_z', \widetilde{\mathbf{q}}) = \sum_{k_z'} W(k_z, k_z'). \qquad (A1)$$

Using Fermi's golden rule, the rate of a single scattering event can be written as follows where both momentum and energy are conserved and ψ corresponds to the mixed (electron and phonon) states.

$$W(k_z, k'_z, \widetilde{q}) = \frac{2\pi}{\hbar} \left| \langle \psi_{k_z} | H_{eP} | \psi_{k'_z} \rangle \right|^2 \delta(E(k'_z) - E(k_z) \pm \hbar\omega(\widetilde{q})). \delta_{q_z,\ k'_z - k_z}. \quad \text{(A2)}$$

The electron-phonon interaction Hamiltonian for phonons of LA type is given as:

$$H_{eP} = D \sum_{\bar{q}} i |\bar{q}| \sqrt{\frac{\hbar}{2\rho V \omega(\bar{q})}} \left( a_q e^{iq.r} + a_q^\dagger e^{-iq.r} \right). \quad \text{(A3)}$$

Where $a_q$ and $a_q^\dagger$ are annihilation and creation operators. Since the z component of the phonon wave vector ($q_z$) is determined by conservation of momentum i.e. $q_z = k_{z'} - k_z$, the summation over phonon wave vectors spans all transversal components of phonon wave vectors.

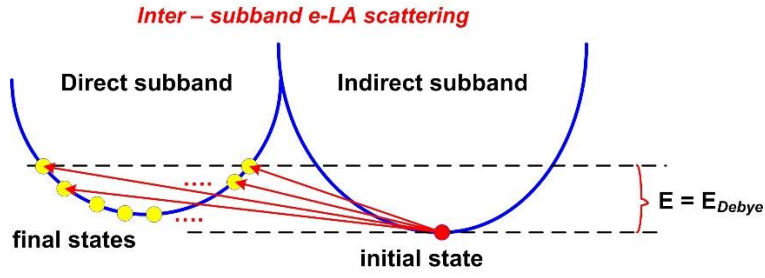

FIG. A.1 Inter-sub band electron-LA phonon scattering events start from $k_z$.

Using the same procedure as discussed in [22] the electron-LA phonon interaction Hamiltonian matrix element is reduced to equation 4.62 in which U stands for Bloch part of the electronic states.

$$\left| \langle U_{k_z} | H_{eP} | U_{k'_z} \rangle \right|^2 = \frac{D^2 \hbar}{2\rho V} \frac{|\bar{q}|^2}{\omega_q} |S(|\bar{q}|)|^2 B_\pm(|\bar{q}|). \quad \text{(A4)}$$

B stands for the Bose-Einstein factor of phonons and it is:

$$B_\pm(|\bar{q}|) = \begin{cases} 1/(e^{\frac{\hbar\omega_q}{K_B T}} - 1) \ for \ absorption \\ 1 + 1/(e^{\frac{\hbar\omega_q}{K_B T}} - 1) \ for \ emission \end{cases} \quad \text{(A5)}$$

With transversal (radial) and longitudinal components of phonon wave vectors ($q_t$ and $q_z$), the absolute value of phonon wave vector can be written as $|\bar{q}| = \sqrt{q_t^2 + q_z^2}$. The structure factor S is defined as follows where $m$ and $m'$ are index of atoms in one unit cell. $k_1$ and $k_2$ denote the electron wave vector (momentum) of two different states and $r_m$ is the coordinate of $m$'th atom.

$$S(|\overline{\pmb{q}}|) = \sum_m |C_{1m}(k_1)|^2 |C_{2m}(k_2)|^2$$

$$+ \sum_{m,m',m\neq m'} C_{1m}(k_1)\,C_{2m}^*(k_2)\,C_{2m'}^*(k_2)\,C_{1m'}(k_1)\,e^{-i\overline{q}.(r_{m'}-r_m)}. \qquad \text{(A6)}$$

It is assumed that there is no overlap between the atomic orbitals of two neighboring unit cells i.e. the orbitals which belong to the same atom can have nonzero overlapping (interaction). The coefficients, $C_{1m}$, are the elements of the nanowire Eigen state ($N_{orbit} \times 1$ vector) at $k_l$ and as explained before they contain 10 numbers corresponding to orbitals of Si atom with index ($m$). Inserting equation (A6) in (A4) and using the result in equation (A1) yields:

$$W_{k_z} = \frac{V}{(2\pi)^3}.\frac{2\pi}{\hbar} \iiint |\overline{q}|^2 \frac{D^2\,\hbar}{2\rho V \omega(\overline{\pmb{q}})} B_{\pm}(|\overline{\pmb{q}}|).|S(|\overline{\pmb{q}}|)|^2$$

$$\times \delta(E(k_z') - E(k_z) \pm \hbar\omega(\overline{\pmb{q}})).\delta_{k_z,k_z'\pm q_z}.q_t\,dq_t\,d\varphi\,dq_z \qquad \text{(A7)}$$

Using the linear dispersion of LA phonon i.e. ($\omega = v_s|\mathbf{q}|$) and after simplification with the aid of Dirac's Delta function properties we have:

$$W_{k_z} = \frac{D^2}{8\pi^2\rho v_s} \iint q_t \sqrt{q_t^2+q_z^2}\,B_{\pm}(|\overline{\pmb{q}}|).\Phi(q_t,q_z) \times \delta(E(k_z')-E(k_z)\pm\hbar\omega(\overline{\pmb{q}})).\delta_{k_z,k_z'\pm q_z}.dq_t\,dq_z$$

$$\qquad \text{(A8)}$$

In which $\Phi$ is defined as equation (A9) and $\varphi$ is the argument of transversal phonon wave vector in polar coordinate.

$$\Phi(q_t,q_z) = \int_0^{2\pi} |S(q_t,q_z,\varphi)|^2 d\varphi \qquad \text{(A9)}$$

Replacing $|\overline{\pmb{q}}|$ with $\sqrt{q_t^2+q_z^2}$ and using Krönecker's delta which imposes $q_z = k_z' - k_z$ i.e.,

$$E(k_z') - E(k_z) = E(k_z \pm q_z) - E(k_z) = \Delta E_{kk'} \qquad \text{(A10)}$$

From (A.8) we obtain:

$$W_{k_z} = \frac{D^2}{8\pi^2\rho v_s}\frac{1}{\hbar v_s} \iint q_t \sqrt{q_t^2+q_z^2}.B_{\pm}(|\overline{\pmb{q}}|).\Phi(q_t,q_z)\,.\,\delta\left(\frac{\Delta E_{uv}}{\hbar v_s}\pm\sqrt{q_t^2+q_z^2}\right)dq_t\,dq_z \qquad \text{(A11)}$$

The integration over $q_t$ can be more simplified using the properties of Dirac's Delta function.

$$W_{k_z} = \frac{D^2}{8\pi^2\rho\hbar^3 v_s^3} \int \Delta E_{kk'}{}^2 \left( B_{\pm}\left(\left|\pm\frac{\Delta E_{kk'}}{\hbar v_s}\right|\right)\right)\Phi\left(q_t = \sqrt{(\frac{\Delta E_{kk'}}{\hbar v_s})^2 - q_z^2},q_z\right)dq_z \qquad \text{(A12)}$$

In which the integration over $q_z$ is converted to a discrete summation over grid points along the 1D BZ. Rewriting equation (A12) and (A1) reveals how it is possible to single out individual rate, $W(k_z, k_{z'})$, between a pair of given states. Recalling that $\Delta q_z = \Delta k_z'$ we can write:

$$W_{k_z} = \sum_{k_z'} W(k_z,k_z') = \sum_{k_z'} \frac{D^2}{8\pi^2\rho\hbar^3 v_s^3}\Delta E_{kk'}{}^2 B_{\pm}\left(\left|\pm\frac{\Delta E_{kk'}}{\hbar v_s}\right|\right)\Phi\left(q_t = \sqrt{(\frac{\Delta E_{kk'}}{\hbar v_s})^2 - q_z^2},q_z\right)\Delta k_z' \qquad \text{(A13)}$$

The summand in Equation (A13) is the same as what is given in equation 3 of section II. For the case of electron (hole)-LO phonon scattering, the same procedure as above is repeated. This time the carrier-LO phonon interaction Hamiltonian of Deformation potential type is used. Readers are encouraged to look at the supplementary material of [5] as well as [21-22].

**Acknowledgments:** This paper was made possible by a NPRP Grant 5-968-2-403 from the Qatar National Research Fund (a member of Qatar Foundation). The statements made herein are solely the responsibility of the authors.